# Observation of Ultrafast Interfacial Exciton Formation and Recombination in Graphene/$MoS_2$ Heterostructure

Yuqing Zou[1], Qiu-Shi Ma[2], Zeyu Zhang[3], Ruihua Pu[4], Wenjie Zhang[1], Peng Suo[1], Jiaming Chen[1], Di Li[1], Hong Ma[5], Xian Lin[1], Yuxin Leng,[3] Weimin Liu[4], Juan Du[3] and Guohong Ma[1*]

[1]Shanghai Frontiers Science Center of Quantum and Superconducting Matter States, Department of Physics, Shanghai University, Shanghai 200444, China

[2]Department of Chemistry, Marquette University, Milwaukee 53233, USA

[3]State Key Laboratory of High Field Laser Physics and CAS Center for Excellence in Ultra-intense Laser Science, Shanghai Institute of Optics and Fine Mechanics, Chinese Academy of Sciences (CAS), Shanghai 201800, China

[4]School of Physical Science and Technology, ShanghaiTech University, Shanghai 201210, China

[5]School of Physics and Electronics, Shandong Normal University, Jinan 250014, China

***Email address: phymagh@shu.edu.cn**

**ABSTRACT**: Heterostructures constructed from graphene and transition metal dichalcogenides have established a new platform for optoelectronic applications. Although massive studies have been done on optical response of the heterostructures in recent years, fundamental questions concerning the dynamical charge transfer and subsequent relaxation are still under debate. Hereby, we combined time-resolved terahertz (THz) spectroscopy along with transient absorption (TA) spectroscopy to revisit the interlayer non-equilibrium carrier dynamics in largely lateral size Gr/$MoS_2$ heterostructure fabricated with chemical vapor deposition (CVD) method. Our experimental results reveal that, with photon-energy below the A-exciton of $MoS_2$ monolayer, hot electrons transfer from graphene to $MoS_2$ takes place in time scale of less than 0.5 ps, resulting in ultrafast formation of interfacial exciton in the heterostructure, subsequently, recombination relaxation of the interfacial exciton occurs in time scale of ~18 ps. A new model considering carrier heating and photogating effect in graphene is proposed to estimate the amount of carrier transfer in the heterostructure, which shows a good agreement with experimental result. Moreover, when the photon-energy is on-resonance with the A-exciton of $MoS_2$, photogenerated holes in $MoS_2$ are transferred to graphene layer within 0.5 ps, leading to the formation of interfacial exciton, the subsequent photoconductivity (PC) relaxation of graphene and bleaching recovery of A-exciton in $MoS_2$ take place around ~10 ps time scale, ascribing to the interfacial exciton recombination. The faster recombination time of interfacial exciton with on-resonance excitation could come from the reduced interface barrier caused by bandgap renormalization effect. Our study provides deep insight into the understanding of interfacial charge transfer as well as the relaxation dynamics in graphene-based heterostructures, which are promising for the applications of graphene-based optoelectronic devices.

## 1.Introduction

Atomically-thin transition metal dichalcogenides (TMDs) is a type of two-dimensional material with excellent properties. It had been widely studied for its visible band gap energy, valley dependent optical selection rule, phase transition and superconductivity behavior.[1-3] In recent years, the study of two-dimensional materials has advanced to artificially superimposing two or more different atomic layers on top of each other to form a new Van der Waals (VdW) heterostructure, including TMDs/TMDs[4-7] and Graphene (Gr)/TMDs.[8-12] The VdW heterostructure is no longer limited by the lattice matching condition and atomic interdiffusion that required in traditional heterostructures, providing a unique platform for the development of optoelectronic devices. Recently, great attention has been focused on a heterostructure composed of graphene and monolayer TMDs, in which graphene processes high carrier mobility and electrical conductivity, while TMDs have high absorption coefficient in visible and near infrared regime, the Gr/TMDs heterostructures can benefit the advantages of both graphene and TMDs which have shown great promises for optoelectronic devices, such as solar cell[13-14], phototransitor[13, 15] and photodectector in far IR and THz regimes,[13, 16-18] *etc*.

Recently, lots of studies have been carried out to unveil the mechanisms of energy and/or charge transfer in the Gr/TMDs heterostructures as well as the subsequent relaxation following photoexcitation. Massicotte *et al*. proposed a photothermal electron emission mechanism with below band gap photoexcitation, and they demonstrated that the hot electrons above the interface energy barrier could be injected into TMDs.[19] Subsequently, Fu *et al*. have verified this in experiments.[20] Using ultrafast microscopy, Yuan *et. al*. demonstrated that both interlayer charger-transfer transition and hot carrier injection promote electrons from graphene to $WS_2$ in Gr/$WS_2$ heterostructure with below bandgap excitation.[8] Instead, Chen *et al*. proposed that hot electrons can be efficiently collected from graphene before electron-hole thermalization.[21] Moreover, the interfacial charge recombination time also varies greatly, Yuan *et al* reported very fast charge recombination time of ~1 ps,[8] and this time constant is very close to the lifetime of charge-separated transient state obtained by time- and angle-resolved photoemission spectroscopy reported by Aeschlimann *et al*.[22] Whereas charge separation of more than 1 ns has been reported in.[20] Therefore, the photocarrier dynamics in Gr/TMDs interface, such as charge transfer occurs thermally or non-thermally, timescales of electron/hole transfer and recombination, existence and assignment of charge-separated transient state, as well as evaluation of transfer efficiency *etc*., remains elusive, and need to be further explored and analyzed. In this article, aiming to independently track the charge carrier dynamics in both graphene and TMDs, time-resolved THz spectroscopy and TA spectroscopy are employed to investigate the charge transfer and the subsequent relaxation process of Gr/$MoS_2$ heterostructure, including the identification of electron and hole transfer, assignment and relaxation of charge-separated transient state, and assessment of the numbers of carrier transfer. Our experimental results show that upon photoexcitation below the A-exciton resonance of $MoS_2$, the hot electrons in graphene can cross the energy barrier and transfer to $MoS_2$ efficiently within time scale of less than 0.5 ps, resulting in formation of interfacial exciton with electron in conduction band edge of $MoS_2$ and hole remained in graphene. The lifetime of the interfacial exciton is determined to be 18 ps. Notably, a combination of theoretical and experimental evaluation of the number of carrier transfers suggests that hot electrons above the energy barrier may not all be transferred to $MoS_2$ and some of them rapidly return to graphene. Moreover, our data support that when photoexcitation energy is near resonant with the A-exciton of $MoS_2$, direct hole transfer from $MoS_2$ to valence band of graphene occurs, leading to the formation of interfacial exciton in the heterostructure, and the interfacial exciton relaxation time is found to be ~10 ps, showing almost 2-fold faster than that in below A-exciton excitation. This suggests that the resonance excitation could modify Gr/$MoS_2$ interfacial barrier due to bandgap renormalization effect. Our experimental study demonstrates that both below and above A-exciton excitation promotes ultrafast formation of interfacial exciton that relaxes in the time scale of 10-20 ps depending on excitation condition. This finding provides insight to further optimize performance of graphene-based optoelectronic devices.

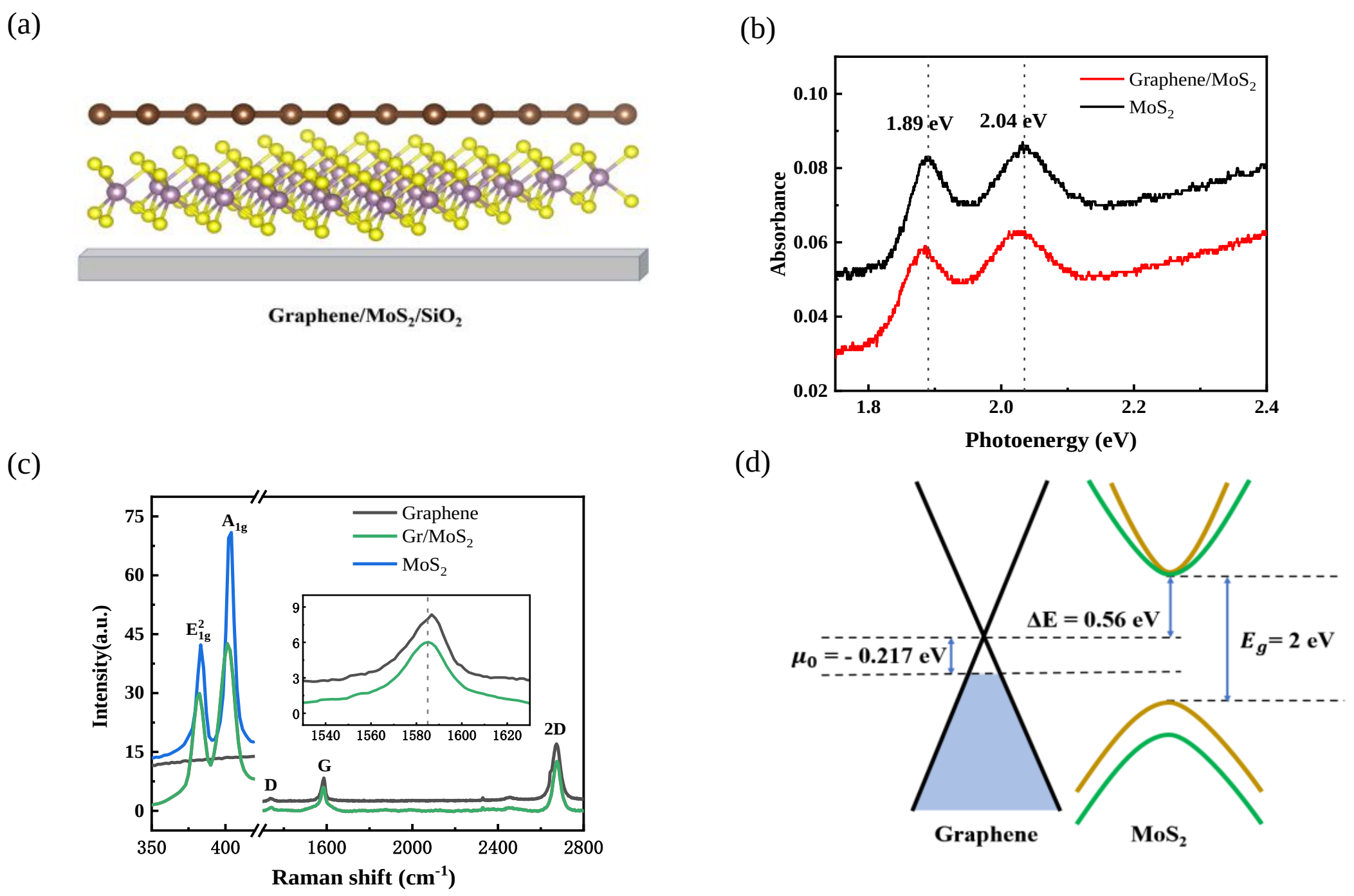

Figure 1. Fabrication and characterization of samples. (a) Schematics of Gr/$MoS_2$ heterostructure. (b) UV-visible absorption spectra for $MoS_2$ (black) and Gr/$MoS_2$ heterostructure (red). (c) Raman spectra of graphene (black), $MoS_2$ (blue) and Gr/$MoS_2$ (green). (d) Schematic diagram of energy band structure for Gr/$MoS_2$ heterostructure.

## 2. Results and Discussion

**2.1 Characterization of samples.** Graphene and $MoS_2$ monolayer thin films were grown by CVD method (provided by Sixcarbon Tech, Shenzhen, China). The structure of Gr/$MoS_2$ heterostructure is illustrated schematically in **Figure 1**a, and the composite film was fabricated by transferring a graphene onto monolayer $MoS_2$ grown on a 1.0 mm-thick $SiO_2$ substrate. The Gr/$MoS_2$ heterostructure along with individual graphene and $MoS_2$ monolayer on silica substrate were characterized by UV-visible absorption spectroscopy and Raman spectroscopy before conducting transient optics and THz experimental measurement. The absorption spectra of $MoS_2$ and Gr/$MoS_2$ are shown in **Figure 1**b. The significant absorption peak at ~1.89 eV and 2.04 eV correspond to A- and B-exciton of monolayer $MoS_2$, respectively, which are in excellent agreement with the previous literatures.[23-24] Peaks A and B are derived from exciton transitions in the spin-split valence bands at K point in Brillouin region in monolayer $MoS_2$. It is noted that the A- and B-exciton peaks in Gr/$MoS_2$ both are red-shifted by amount ~60 meV comparing to the monolayer $MoS_2$ due to the coupling effect between graphene and $MoS_2$ in the heterostructure. The Raman spectra taken on samples are shown in **Figure 1**c. There are three characteristic peaks of graphene layer in Gr/$MoS_2$: G-, D- and 2D-band are located at ~1585 $cm^{-1}$, ~1342 $cm^{-1}$ and ~2675 $cm^{-1}$, respectively. The strength ratio of 2D-band to G-band is about 2, suggesting that graphene layer is a high quality monolayer.[25-26] Two characteristic peaks of $MoS_2$ are located at ~ 382 $cm^{-1}$ ($E_{1g}^2$) and ~ 401 $cm^{-1}$ ($A_{1g}$), respectively under graphene covered, and frequency difference between $A_{1g}$ and $E_{1g}^2$ modes is about 19 $cm^{-1}$, which further indicates the single layer nature of $MoS_2$ in the heterostructure.[27-29] After being covered by graphene, the $A_{1g}$ peak is down-shifted by 1 $cm^{-1}$ with a little wider peak width compared with the monolayer $MoS_2$, indicating that the $MoS_2$ layer in the heterostructure becomes more n-doped.[30-31] According to the G-band position ~1585 $cm^{-1}$ of Gr/$MoS_2$, we can estimate that the Fermi level ($E_F$) of graphene is about 0.12 eV below Dirac point,[31-32] and the chemical potential at room temperature is about 0.218 eV below Dirac point from $\mu = E_F^2/4\ln(2)K_B T$ .[33] The value of the quasiparticle band gap of $MoS_2$ is 2.0 eV,[34] and the energy difference between graphene Dirac point and $MoS_2$ minimum conduction band is 0.56 eV,[35-37] as illustrated in **Figure 1**d. Therefore, we estimate the energy barrier for electron, $\Delta E_e$ (hole, $\Delta E_h$) between chemical potential of graphene and minimum (maximum) of conduction (valence) band of $MoS_2$ is ~0.778 eV (1.222 eV).

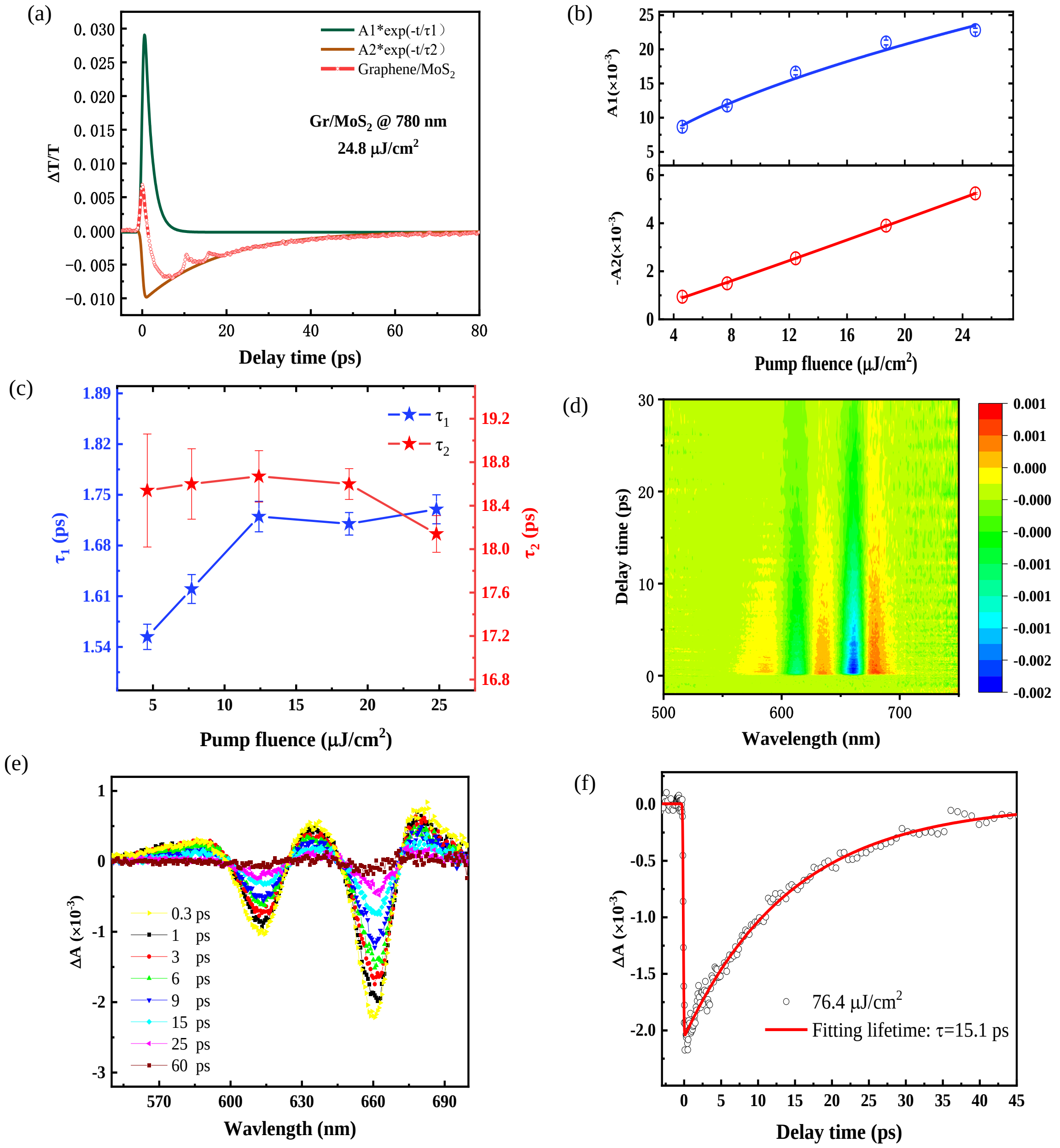

Figure 2. Transient THz transmission and TA spectra for Gr/$MoS_2$ with below A-exciton photoexcitation of 1.59 eV (780 nm). (a) Transient THz transmission of Gr/$MoS_2$ (red), green and yellow solid lines are biexponential fitting convoluted with instrumental response function. The pump fluence was fixed at 24.8 $\mu J\ cm^{-2}$. (b) and (c) pump fluence dependent fitting weights ($A_1$and $A_2$) and lifetimes ($\tau_1$ and $\tau_2$) of (a). (d) Two-dimensional plots of TA color map for Gr/$MoS_2$ following 1.59 eV excitation with pump fluence of 76.4 $\mu J\ cm^{-2}$. (e) Representative TA spectra probed at several selected delay times. (f) Transient trace of photobleaching at 660 nm in (e), the red solid line is single exponential fitting with fitting lifetime of 15.1 ps.

Time-resolved THz spectroscopy is sensitive to conductivity change of free carriers, which has been demonstrated to be a sensitive tool to probe carrier dynamics in graphene. On the other hand, TA spectroscopy in visible region is a good spectroscopic tool to sample the exciton dynamics in TMDs like $MoS_2$ monolayer. Combined two spectroscopic tools together can provide a clear picture of the interfacial charge transfer. Section S1 in Supporting Information (SI) schematically shows the experimental configurations for transient THz spectroscopy and TA spectroscopy. Here, we tune the pump energy to excite the graphene layer only ($h\nu$<1.89 eV, namely below A-exciton excitation) in Gr/$MoS_2$ or both layers together ($h\nu$ >1.89 eV, on resonance or above A-excitation) to investigate the interfacial charge transfer (CT) process. The pump-induced PC ($\Delta\sigma$) can be obtained by monitoring the transient THz transmission ($\Delta \mathrm{T} = \mathrm{T}_{pump} - T_0$) according to Eq. (1),[38] in which $n$=1.95 is refractive index of $SiO_2$ substrate, $Z_0$=377 is the impedance of free-space, and $T_{pump}$ and $T_0$ denote the transmitted THz electric field with and without optical excitation, respectively.

$$\Delta\sigma = -\frac{(1+n)}{Z_0 d}\,\frac{\Delta\mathrm{T}(t)}{T_0(t)} \qquad (1)$$

**2.2 Photocarrier dynamics in Gr/$MoS_2$ with below A-exciton photoexcitation.** Firstly, we have tested the transient response of individual graphene, $MoS_2$ and Gr/$MoS_2$ under identical photoexcitation of 1.59 eV. The PC of graphene is negative, and the lifetime obtained by single exponential fitting is about 1~2 ps depending on pump fluence. Since the photon energy of 1.59 eV is lower than A-exciton of $MoS_2$, no signal is observed in $MoS_2$ (see S2 in SI). **Figure 2**a displays transient THz transmission of Gr/$MoS_2$ with pump fluence ~24.8 $\mu J$ $cm^{-2}$. It is obviously observed that the PC of Gr/$MoS_2$ changes from negative to positive in a few ps, indicating that carriers in graphene are transferred to $MoS_2$. The two small peaks located around at the delay time of 10 ps and 16 ps are the secondary reflections from the sample as well as the THz emitter ZnTe, respectively. As the magnitude of $\Delta\sigma$ is the production of carrier density change $\Delta n$ and carrier mobility $\mu$, i.e. $\Delta\sigma = \Delta n e \mu$ with e of electron charge. Positive PC may result from a reduction in graphene Fermi energy levels due to electron extraction from graphene and an increase in the free carrier concentration in $MoS_2$ layer. It is well known that the carrier mobility of graphene ($10^3 \sim 10^4$ $cm^2V^{-1}S^{-1}$)[39-40] is 2-order of magnitude higher than that of $MoS_2$ ($20 \sim 300$ $cm^2V^{-1}S^{-1}$),[41-43] therefore, the positive PC after below A-exciton photoexcitation is mainly contributed by graphene layer, i.e. the increase hole concentration of graphene. In contrast to the previously reported charge separation of more than 1 ns at the interface of Gr/$WS_2$ heterostructure,[20] here carrier relaxation time of Gr/$MoS_2$ determined from transient THz spectroscopy is about 18 ps. In order to further illustrate that photoexcitation induced hot carrier transfer rather than non-thermal charge transfer. We performed OPTP measurement on Gr/$MoS_2$ with much lower pumping photon-energy of 0.95 eV and 0.775 eV. It is clear photoelectron energy in graphene ($E = \frac{1}{2}h\nu$) is lower than the electron interfacial barrier ($\Delta E = 0.778$ $eV$) under photoexcitation at 1300 nm ($h\nu = 0.95$ $eV$) and 1600 nm ($h\nu = 0.775$ $eV$), not surprisingly, the photoconductivity response in Gr/$MoS_2$ shows similar profile as the case of 780 nm excitation (see section S3 in SI). The transient THz transmission ($\Delta T/T_0$) can be well fitted with biexponential function convoluted with a Gaussian function as shown in Eq. (2),

$$\frac{\Delta T}{T_0}(t) = A_1 \cdot e^{\frac{\omega^2}{\tau_1^2}-\frac{t}{\tau_1}} \cdot erfc\left(\frac{\omega}{\tau_1} - \frac{t}{2\omega}\right) + A_2 \cdot e^{\frac{\omega^2}{\tau_2^2}-\frac{t}{\tau_2}} \cdot erfc\left(\frac{\omega}{\tau_2} - \frac{t}{2\omega}\right) \qquad (2)$$

in which $\tau_1$ and $\tau_2$ are the relaxation time with corresponding amplitude of $A_1$ and $A_2$, respectively, $2\omega$ is the full width at half maximum (FWHM) of THz waveform, and $erfc(x)=1-erf(x)$ is the complementary error function. The biexponential fitting results are displayed in **Figure 2**a (more fitting data are shown in section S6 of SI). It is noted that the transient THz trace shows identical profile for different photoexcitation energy (lower than A-exciton of $MoS_2$), which is displayed in section S4 of SI for comparison. **Figure 2**b and 2c present the fitting weight (2b: $A_1$ and $A_2$) as well as lifetime (2c: $\tau_1$ and $\tau_2$) with respect to pump fluence, respectively. It is clear the weight $A_1$ is positive, corresponding to the negative PC of Gr/$MoS_2$ due to hot electrons scattering of graphene, while the negative $A_2$ corresponds to the positive PC due to the increase of hole concentration in graphene after hot electrons transfer occurs. The fitting index $\alpha$~0.57 is obtained by fitting pump fluence dependence of $A_1$ with function $I^{\alpha}$, where I is pump fluence. Similarly, the pump fluence dependence of $A_2$ give rise to a fitting index $\alpha$ of 0.94, which is notable in contrast to the recently reported superlinear pump fluence dependence of positive photoconductivity.[19,20] Furthermore, our data show a sublinear dependence at high power (section S3 in SI). In this case, we suggest that the thermal carriers crossing the interfacial barrier do not fully transfer to $MoS_2$, and part of the thermal electron return back to graphene layer, resulting in an approximately linear pump fluence dependence. The fitting lifetimes, $\tau_1$ and $\tau_2$, shown in **Figure 2**c correspond to fast and slow components, respectively. The fast lifetime $\tau_1$ increases slightly with pump fluence, which arises from hot electron cooling in graphene, and the slow lifetime $\tau_2$~18 ps does not show obviously fluence dependence, which is believed to arise from interfacial exciton recombination driven by Coulomb attraction.

The consensus on photoinduced ultrafast charge transfer has been achieved in coupled Gr/TMDs heterostructure in literatures,[8,10,14,17,19-22] our THz experimental data here also verify that hot electron injection from graphene to $MoS_2$ with below A-exciton excitation dominates the charge transfer process in the Gr/$MoS_2$ heterostructure. However, the subsequent relaxation has been reported to vary from 1 ps to 1 ns in literatures,[8,10,17,19-22] and the mechanism behind is still controversial. We tentatively proposed that the relaxation of photoconductivity of graphene with typical time constant of 18 ps could arise from the recombination of interfacial exciton. In fact, photoexcitation heats the electrons in graphene, resulting in Fermi-Dirac redistribution of hot electrons. The hot electrons with energy higher than interfacial barrier are transferred to the conduction band of $MoS_2$, which could lead to the formation of interfacial excitons with electrons in $MoS_2$ conduction band and holes in valence band of graphene. As a result, graphene is positively charged and $MoS_2$ is negatively charged, the Coulomb attraction between the electrons and holes drives the occurrence of the interfacial exciton recombination, which takes place in time scale of 18 ps in our

$Gr/MoS_2$ heterostructure.

In order to further consolidate our assumption about the interfacial exciton recombination following photoexcitation, we study the dynamics of A-exciton of $MoS_2$ in Gr/MoS2 heterostructure by TA spectroscopy with below A-exciton excitation. If interfacial excitons are formed after photoexcitation, it is then expected that the electrons relaxation in $MoS_2$ should follow similar relaxation process as that of hole in graphene. **Figure 2**d displays the two-dimensional TA color diagram of $Gr/MoS_2$ with photoexcitation of 1.59 eV, and the pump fluence is fixed at 76.4 $\mu J\ cm^{-2}$. No detectable TA signal was observed for individual graphene and $MoS_2$ (see section S2 in SI), but pronounced photobleaching signal around A-exciton (660 nm) and B-exciton (610 nm) was observed for $Gr/MoS_2$. **Figure 2**e illustrates the TA spectra in the range from 550 nm to 700 nm collected at several delay times, further indicating that electron transfer occurs from graphene to conduction band of $MoS_2$. **Figure 2**f shows the TA traces of $Gr/MoS_2$ probing at 660 nm. A typical relaxation time of 15.1-ps by single exponential fitting represents the main recovery of A-exciton bleaching at 660 nm. Clearly, this 15.1-ps lifetime obtained from the recovery of A-exciton bleaching in $MoS_2$ shows identical lifetime scale observed in PC relaxation in graphene layer from transient THz measurement, indicating electron and hole dissipate their energy with the same rate. And the attenuation lifetime is basically the same with various pump fluence (see **Figure S5**a of section S5 in SI). In addition, the recovery of the bleaching signal at 660 nm (A-exciton) shows almost identical as that at 610 nm (B-exciton) as shown in **Figure S5**b, suggesting the recovery of bleaching signals (at 660 nm and 610 nm) is strongly correlated with the relaxation of electrons in conduction band of $MoS_2$. In short, our experimental results reveal that below A-exciton photoexcitation of the $Gr/MoS_2$ heterostructure leads to ultrafast electron transfer from graphene to $MoS_2$ resulting in formation of interfacial exciton, and the subsequent relaxation comes from the interfacial exciton recombination.

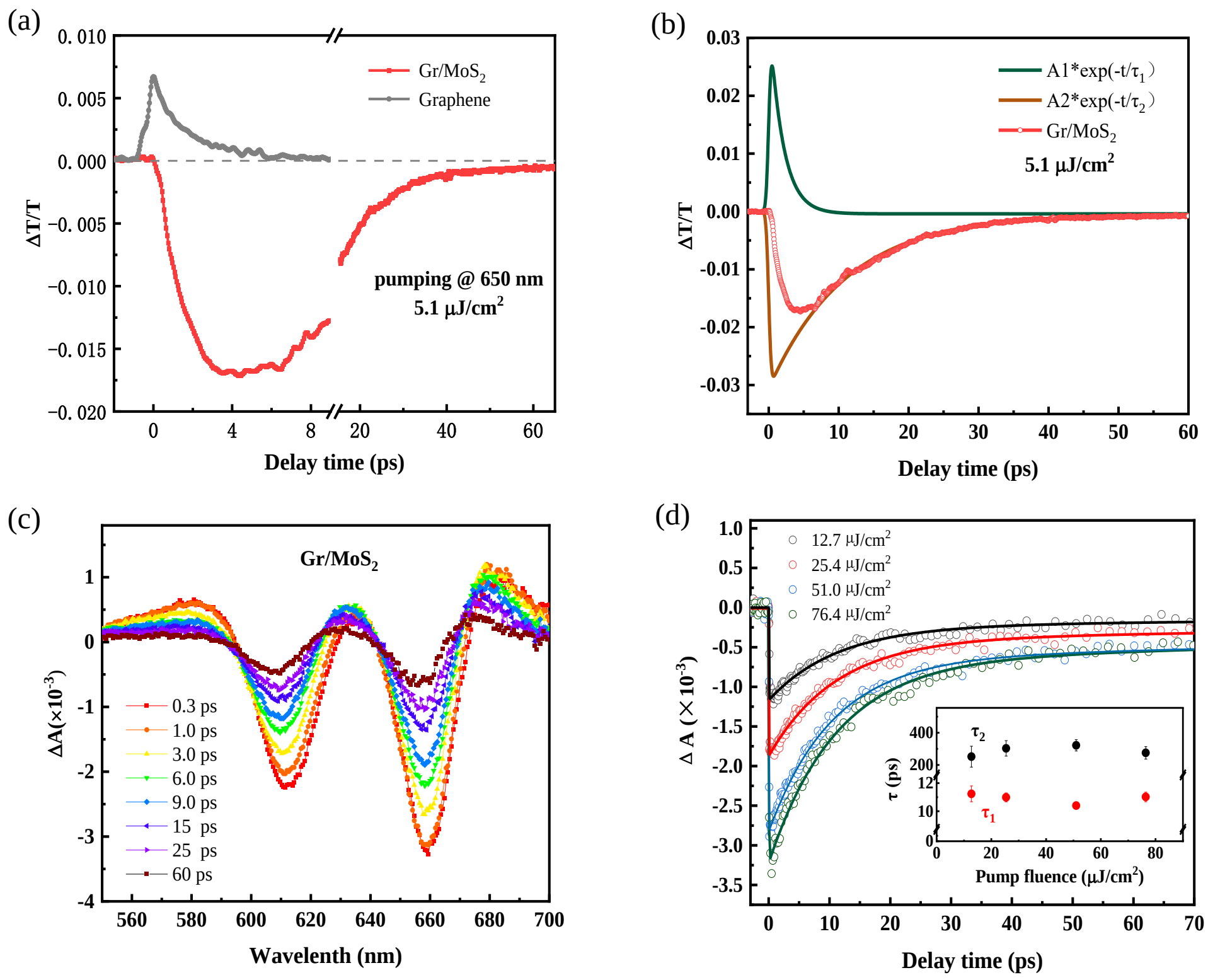

Figure 3. THz spectra and TA spectra of Gr/$MoS_2$ under photoexcitation of 1.9 eV (650 nm). (a) Transient THz transmittance of graphene (gray) and Gr/$MoS_2$ (red), with pump fluence was fixed at 5.1 $\mu J$ $cm^{-2}$. (b) Convolution fitting of double exponential attenuation function with the experimental response function. The red line is experimental data, and the green and yellow lines are the fitting curves. (c) Representative TA spectra of Gr/$MoS_2$ are detected at several delay times. (d) Transient carrier dynamics monitored at 660 nm (A-exciton resonance in heterostructure) at different pump fluences. The solid line is a double exponential function fitting, and the recovery time of A-exciton bleaching obtained is shown in the inset.

**2.3 Photocarrier dynamics in Gr/$MoS_2$ with above A-exciton photoexcitation.** In order to further demonstrate the ultrafast charge-transfer and formation of interfacial exciton, we have investigated the photocarrier dynamics of Gr/$MoS_2$ heterostructure under photoexcitation of 650 nm (1.9 eV), near resonance with the A-exciton. We also have tested THz response of graphene and $MoS_2$ monolayer under identical condition, and **Figure 3**a plots the experimental data of both graphene and heterostructure together for comparison. As expected, graphene exhibits negative photoconductivity and follows a single exponential decay with a lifetime of 1~2 ps. While the $MoS_2$ monolayer shows no detectable photoinduced THz signal. Since the photon energy of pump pulse is near on-resonance with A-exciton of $MoS_2$, both graphene and $MoS_2$ in Gr/$MoS_2$ can be excited. By comparing the THz response under 1.9 eV excitation (**Figure 3**a) and 1.59 eV (**Figure 2**a), the negative PC of Gr/$MoS_2$ under 1.9 eV excitation is less prominent, while the positive PC signal becomes more pronounced than that under 1.59 eV excitation. Although weakly negative PC signal observed in **Figure 3**a, we can still use biexponential function to fit the experimental data, which is presented in **Figure 3**b. It is clear the contribution of negative PC from graphene shows a decay lifetime of 1~2 ps, which is very similar to that with 1.59 eV excitation. However, the positive photoconductivity with much stronger intensity shows a relaxation time of ~10 ps. Considering the large optical absorption of $MoS_2$ around A-exciton and large energy difference between the Fermi level of graphene and maximum valence band of $MoS_2$ in the Gr/$MoS_2$. The photogenerated holes on top valence band of $MoS_2$ can be injected into valance band of graphene efficiently, leading to the decrease of graphene Fermi energy level significantly, which results in the pronounced positive THz PC in Gr/$MoS_2$ upon 650 nm excitation. As a result, interfacial excitons are formed with holes residing in the valence band of graphene and electrons remaining in conduction band of $MoS_2$. The subsequent relaxation of THz PC comes from the interfacial exciton recombination. By recalling the 18 ps hole relaxation time of graphene at 1.59 eV excitation as shown in **Figure 2**c, the relaxation life is accelerated (~10 ps) under 1.9 eV excitation, which is attributed to a change of the interface potential between graphene and $MoS_2$ under resonant excitation of A-exciton. With the on-resonance excitation of A-exciton of $MoS_2$, photogenerated electrons and holes are respectively populated in the conduction and valence bands before charge transfer takes place. Subsequently, interfacial excitons in Gr/$MoS_2$ are formed with holes are transferred to valence band of graphene, and at the same time, intralayer excitons (with remaining electrons in conduction band and holes in valence band of $MoS_2$) are also formed via electron-hole coulomb interaction. The existence of intralayer exciton gives rise to the bandgap renormalization of $MoS_2$,[35,45,50] which could modify the interface barrier of Gr/$MoS_2$, resulting in fast recombination time at 1.9 eV excitation. **Figure** S6 in SI presents the detailed comparisons for below and above A-exciton excitation with various pump fluences along with biexponential fittings.

TA spectroscopy with above A-exciton excitation is also conducted to probe the A-and B-exciton dynamics of $MoS_2$ layer in the heterostructure. **Figure 3**c shows the TA spectra of Gr/$MoS_2$ collected at different delay times under pump fluence of 76.4 μJ $cm^{-2}$ upon excitation of 650 nm, and pronounced photobleaching signals around A- and B-exciton peaks are clearly seen, from which we can conclude that photoexcitation at 650 nm leads to holes transfer occurs from $MoS_2$ to graphene rather than electron transfer.[24,44] If electrons transfer from $MoS_2$ to graphene occurs after 650-nm excitation, and holes remains on the top valence band of $MoS_2$, then only bleaching signal around A-exciton is expected, and no bleaching signal around B-exciton can be observed according to Pauli exclusion principle, this is because both the conduction and valence bands corresponding to B-exciton are not occupied with the assumption of electron transfer from $MoS_2$ to graphene taking place after photoexcitation. In conjunction with THz experimental data given in **Figure 3**a, we conclude that above A-exciton excitation triggers the ultrafast hole transfer occurs from valence band of $MoS_2$ to that of graphene, which leads to the increase of THz photoconductivity and presence of bleaching signal around both A- and B-exciton. **Figure 3**d presents the dynamical relaxation monitoring at 660 nm with several pump fluences, and the dynamics can be well fitted with biexponential function, which produces a fast lifetime of ~11 ps with amplitude of over 75% and a slow one of 250 ~ 300 ps with amplitude of ~25%. The fitting results are shown in the inset in **Figure 3**d, where the short lifetime (~11 ps) is basically consistent with the hole relaxation time (10 ps) of graphene in **Figure 3**b. Another feature of TA spectra in Gr/$MoS_2$ heterostructure is the appearance of hundreds of ps long-lived signal, which has been also observed in B-exciton bleaching as well as the monolayer $MoS_2$ under resonant A-exciton excitation (see the details in Section S7 in SI), we attribute the long lifetime to the relaxation process of the intralayer exciton of $MoS_2$. As a matter of fact, under near resonant excitation of A-exciton, most of photogenerated holes on the top valence band of $MoS_2$ are transferred to graphene, a small portion of holes are still remained on the valence band of $MoS_2$ layer, leading to formation of intralayer exciton in $MoS_2$ monolayer. The large binding energy of newly formed intralayer exciton could cause the bandgap renormalization effect (BRE) of $MoS_2$, as a consequence, the TA spectra shifts with the delay time as shown in **Figure 3**c. More pronounced spectra shift observed in $MoS_2$ monolayer further indicates that bandgap renormalization is more significant than that in Gr/$MoS_2$ heterostructure (see more details in Section S7 of SI). This is because the most of holes are transferred to graphene in the Gr/$MoS_2$ heterostructure, and the population of intralayer exciton in the heterostructure is much less than that in $MoS_2$ monolayer under identical excitation. In addition, the intralayer exciton dissipates the energy by electron-hole recombination, as the case in monolayer $MoS_2$, showing lifetime in the time scale of hundreds of ps. As the intralayer exciton in $MoS_2$ shows almost zero conductivity, and THz radiation is almost blind to the intralayer exciton in $MoS_2$, therefore THz spectroscopy is only sensitive to interfacial exciton in Gr/$MoS_2$ due to the largely different conductivity of holes in graphene and electrons in $MoS_2$. In summary, with photoexcitation of above A-exciton of $MoS_2$, holes of $MoS_2$ are injected into the valence band of graphene, resulting in formation of interfacial exciton with typical lifetime of 10 ps.

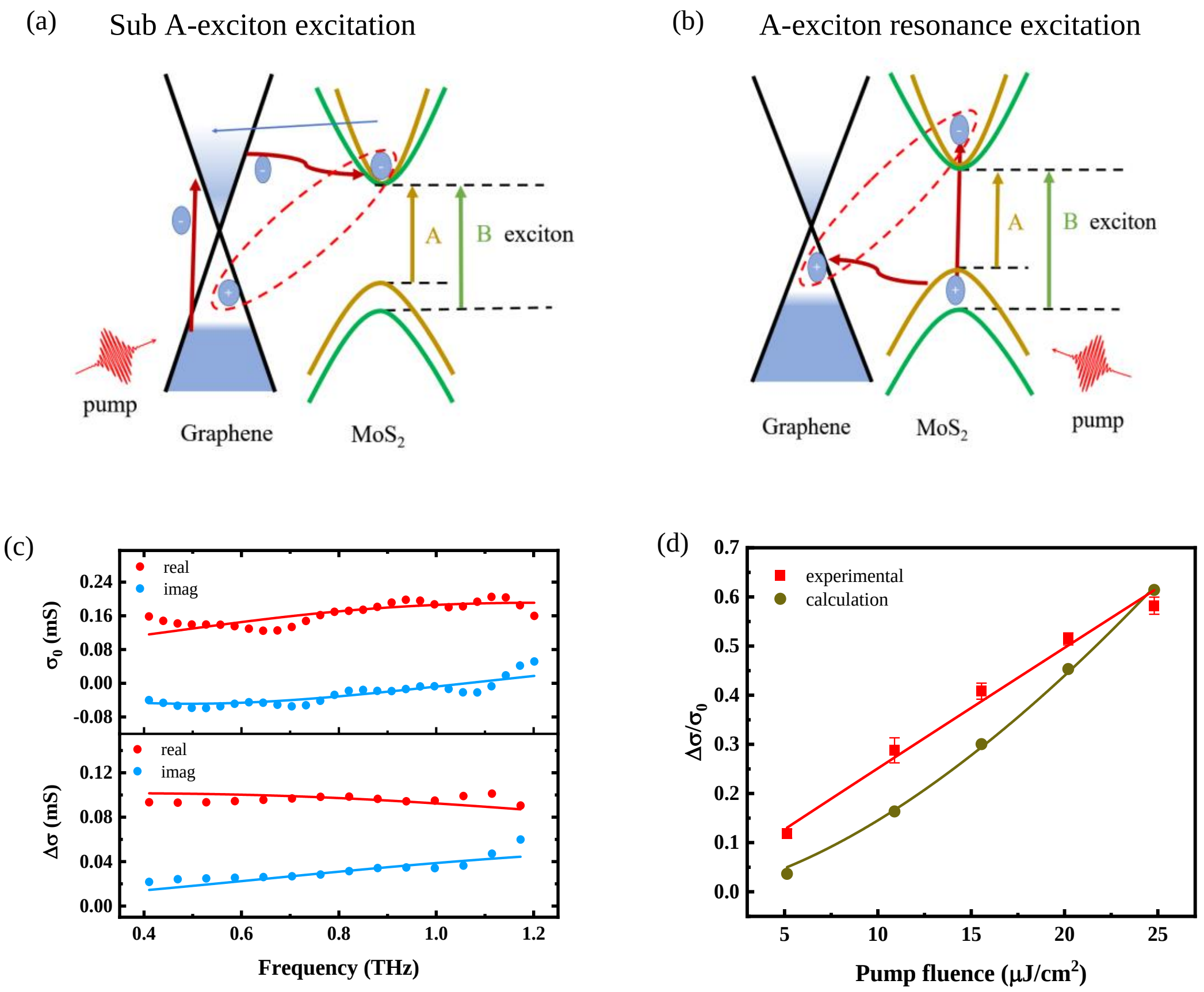

Figure 4. Schematic diagram of Gr/$MoS_2$ interfacial charge transfer and evaluation of the number of carrier transfer. (a) Some of hot electron crossing energy barrier in graphene be injected into conduction band of $MoS_2$ under below A-exciton excitation, the transferred electrons in $MoS_2$ are combined with holes in graphene to form interfacial exciton via the Coulomb interaction. (b) Holes in $MoS_2$ are transferred directly to graphene valence band with above A-exciton excitation, the transferred holes in graphene are combined with the electrons in $MoS_2$ to form interfacial excitons. (c) The upper panel shows the complex static conductivity of Gr/$MoS_2$ without photoexcitation; the lower panel shows the complex photoconductivity of Gr/$MoS_2$ obtained at delay time of 5 ps and a fixed fluence of 24.8 μJ $cm^{-2}$. Solid lines represent the Drude-Smith fits. (d) Pump fluence dependence of the number of carriers are transferred. The red squares are experimental evaluation, which is most almost linearly related to pump fluence with $\Delta\sigma/\sigma_0 \sim I^{0.94}$, and the green circles are the theoretical calculation, which shows superlinearly pump fluence dependence with $\Delta\sigma/\sigma_0 \sim I^{1.6}$.

**2.4 Evaluation of the number of carrier transfer occurs.** As above A-exciton excitation can thermalize the carriers in graphene and populate carrier distribution in $MoS_2$, the evaluation of charge transfer for this case is much more complex than the case below A-exciton excitation. Here, we only give an evaluation of the number of carrier transfer occurs with below A-exciton excitation. It has been proved above that upon excitation below the A-exciton of $MoS_2$, thermalized electrons of graphene can be injected across the energy barrier into $MoS_2$, which leads to the formation of interfacial excitons. However, it is not clear how much carrier transfer actually occurs. It is worth noting that the carrier transfer route and efficiency have been analyzed in previous reports.[20-21] Chen *et al.* proposed that hot electrons can be efficiently extracted from graphene before the electron-hole has been thermalized, with injection quantum yields as high as 50%.[21] Fu *et al.* proposed a photo-thermionic emission scheme, and they defined electron transfer (ET) efficiency as $\eta = N_{ET}(h\nu)/N_{abs}$, in which $N_{ET}(h\nu)$ is the amount of electrons injected from graphene into $MoS_2$ when the excitation energy is $h\nu$, and $N_{abs}$ represents the absorbed photon density.[20] The photocurrent caused by photo-thermionic emission is expected to be superlinear with the excitation power.[19,46] Interestingly, our experimental results show that the photoinduced conductivity is nearly linear with pump fluence. Here, we propose a new model to estimate the number of carrier transfer. As we have described above, the extraction of graphene electron will lead to the reduction of Fermi energy level, thus forming positive photoconductivity as shown in **Figure 2**a. The amount of transferred carriers can be obtained by calculating the change of Fermi energy level of graphene layer in Gr/$MoS_2$ before and after electron transfer takes place. As a consequence, the ratio of the change in photoconductivity ($\Delta\sigma$) following photoexcitation to the static conductivity ($\sigma_0$) without photoexcitation can be used to evaluate the number of transferred carriers. For p-type graphene, the initial hole concentration ($n_0$) of graphene at room temperature can be calculated based on Fermi level and Fermi-Dirac distribution.[47-48] Electron are heated under photoexcitation, the transferred thermal electrons ($\Delta n$) can be obtained according to the barrier height between graphene chemical potential and the minimum conduction band of $MoS_2$ (here, assuming that all the hot electrons above the barrier are transferred to $MoS_2$). The ratio of the two ($\Delta n/n_0$) at different power is shown the green circles in **Figure 4**d, see section S9 of SI for detailed calculation. Experimentally, in order to obtain static conductivity of graphene layer in Gr/$MoS_2$, THz-time domain spectra of Gr/$MoS_2$/$SiO_2$ and $MoS_2$/$SiO_2$ were measured. The transmitted THz intensity was recorded as $E$ for Gr/$MoS_2$/$SiO_2$ and $E_0$ for $MoS_2$/$SiO_2$, respectively. Complex conductivity ($\sigma_0$) of graphene can be obtained from the frequency domain spectra after the Fourier transform, and the real and imaginary parts are then simulated with Drude-Smith model,[38,49] which is shown in the upper panel of **Figure 4**c (see section S8 in SI for details). The lower panel in **Figure 4**c shows the real and imaginary parts of THz PC ($\Delta\sigma$) dispersion for Gr/$MoS_2$ under 780-nm-excitation at pump fluence of 24.8 $\mu J$ $cm^{-2}$. **Figure S8** in SI presents PC dispersion with various pump fluence, from which the pump fluence dependent $\Delta\sigma$ of the heterostructure can be obtained. To avoid the influence of hot electrons, we took the experimental data at time delay of 5 ps, in which the positive THz PC reaches maximum as shown in **Figure 2**a, and the hot electrons in graphene have been cooled down. Considering the flat dispersion in the frequency range from 0.4 to 1.2 THz, we take the average values of the real parts of $\Delta\sigma$ and $\sigma_0$, respectively from 0.4 to 1.2 THz, the pump fluence dependence of $\frac{\Delta\sigma}{\sigma_0} = \frac{\Delta n}{n_0}$ is plotted with red squares in **Figure 4**d. By fitting the data with power exponential function ~$I^{\alpha}$, which produces almost linear dependence with $\alpha$=0.94. This linear fluence dependence is consistent with the maximum positive PC variation shown in **Figure 2**b, but deviates from theoretical prediction of superlinear dependence as shown with green circles, in which power index $\alpha$ is about 1.6. We believe that the hot electrons above the barrier in graphene are not fully injected into $MoS_2$, part of charge carriers may return back to graphene layer, which results in a derivation from the theoretical superlinear pump fluence dependence.

## 3.Conclusion

To summarize, using ultrafast THz time-resolved and TA spectroscopy, we have investigated the interfacial photocarrier dynamics of Gr/$MoS_2$ heterostructure. The experimental results reveal ultrafast charge transfer occurs for both below and above

A-exciton of $MoS_2$ excitation, as well as the formation and recombination of interfacial excitons. With below A-exciton excitation, hot electrons in graphene are transferred to $MoS_2$, leading to the formation of interfacial exciton with electrons in $MoS_2$ and holes in graphene, and the lifetime of the interfacial exciton is determined to be about 18 ps. On the other hand, with above A-exciton excitation, photogenerated holes in $MoS_2$ are transferred to valence band of graphene and electrons are remained in the conduction band of $MoS_2$, as a result, interfacial excitons are formed with typical lifetime of 10 ps. Moreover, by evaluating the number of transferred carriers through theoretical calculation and experimental results, we conclude that not all the hot electrons that above the energy barrier are transferred. Our results provide new insight into understanding the photocarrier dynamics in Gr/TMDs heterostructure, which is promising for potential applications of graphene-based heterostructures in solar cell, photodetectors and the related fields.

## 4. Experimental Section

**Optical pump and THz probe spectroscopy (OPTP): Figure** S1a in SI illustrates the schematic diagram of optical pump and THz probe spectroscopy (OPTP). In our home-built ultrafast OPTP setup, the optical pulses are delivered from a Ti: sapphire amplifier with 120 femtoseconds (fs) duration at central wavelength of 780 nm (1.59 eV) and a repetition rate of 1 kHz. The laser pulse output from the amplifier is divided into three beams: the first two beams are used to generate and detect the THz signal, which constitutes the terahertz time-domain spectroscopy (THz-TDS) and is utilized to measure the static conductivity of the sample, in which the THz emitter and detector are based on a pair of (110)-oriented ZnTe crystals. The third beam acting as pump beam, is used to generate photocarrier in the samples. For the optical pump at other wavelengths, the third beam is guided into optical parametric amplifier (TOPS-prime) to tune the pump wavelength. The pump induced THz transmission change ($\Delta T = T - T_0$) with respect to delay time ($\Delta t$) was recorded, in which T and $T_0$ denote the transmission of the THz electric field peak value with and without pump beam, respectively. The optical pump and THz probe pulse are collinearly polarized with a spot size of 6.5 and 2.0 mm on the surface of sample, respectively. All measurements were conducted in dry nitrogen atmosphere.

**Transient Absorption Spectroscopy**: In order to further explore the dynamics of A-exciton in both Gr/$MoS_2$ heterostructure and $MoS_2$ monolayer, a TA measurements were implemented as shown schematically in **Figure 1**Sb in SI, where the commercial TA spectrometer (HELIOS FIRE, Ultrafast System) is driven by a Ti: sapphire laser (Coherent, Astrella). The laser outputs optical pulses centered at 800 nm with a pulse duration of 35 fs and repetition rate of 1 kHz, which was divided into two beams: one was guided into an optical parametric amplifier to produce 780 nm excitation light, and the other beam was focused on a sapphire slice to generate white light supercontinuum with the pulse duration of 150 fs, which serves as probe beam in the ultrafast system. The change of absorbance (ΔA) induced by the pump optical pulse (ΔA) reflects the photoinduced bleaching ($\Delta A>0$) and/or photoinduced absorption ($\Delta A<0$). During the measurement, samples were placed on a holder and moved within a $1\times1$ $mm^2$ plane perpendicular to the probe light propagation direction to avoid thermal accumulation. All the measurements were carried out at room temperature.

### Supporting Information

Supporting Information is available from the authors

### Acknowledgements

This work is supported by the National Natural Science Foundation of China (Grant, Nos. 92150101, 6173501), and Natural Science Foundation of Shanghai (Grant No. 20ZR1436300)

### Conflict of interest

The authors declare no conflict of interest.

## Supporting Information

### Section S1: Schematics of time-resolved THz spectroscopy and transient absorption Spectroscopy

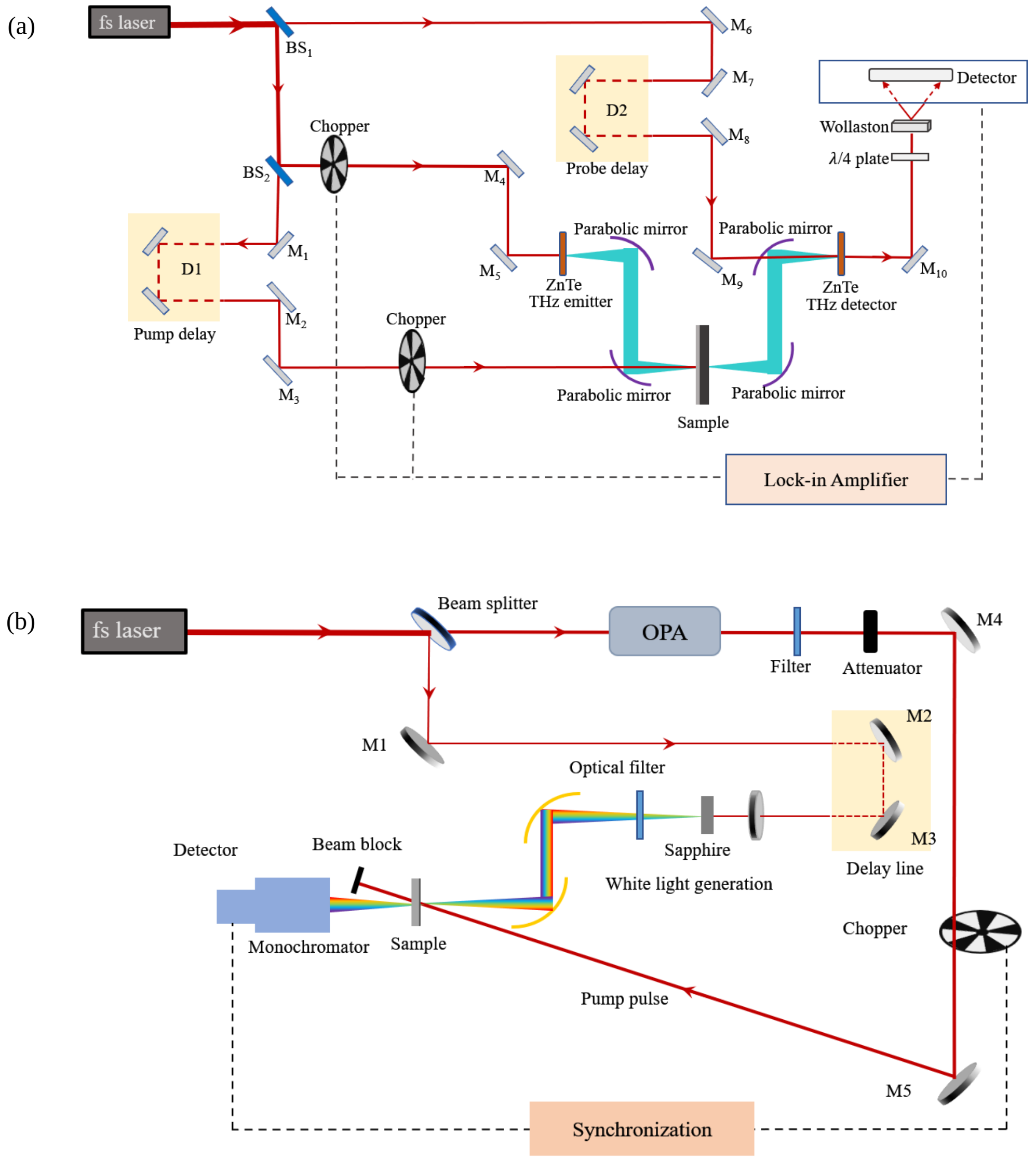

Figure S1. (a) Schematic diagram of optical pump and THz probe spectroscopy for measuring the photoconductivity of Gr/$MoS_2$ heterostructure and graphene monolayer. (b) Schematic diagram for transient absorption spectroscopy for measuring the photocarrier dynamics of Gr/$MoS_2$ heterostructure as well as $MoS_2$ monolayer.

## Section S2: Transient THz and TA spectra of graphene and $MoS_2$ under 780 nm (1.59 eV) photoexcitation.

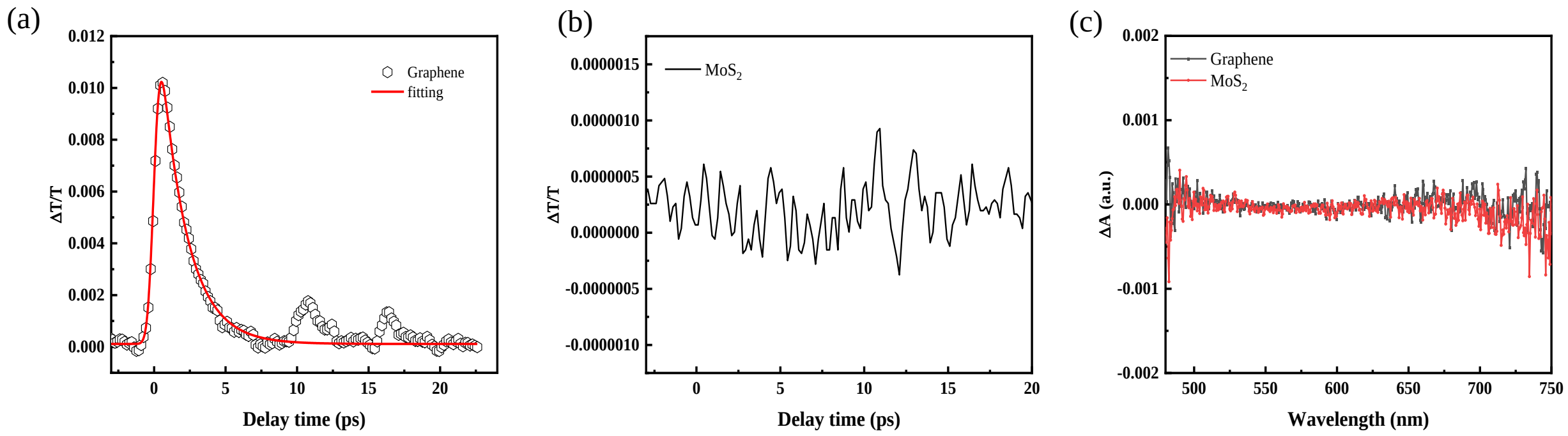

Figure S2. (a) The transient THz transmission of graphene under 780 nm photoexcitation, the positive $\Delta T/T_0$ indicates negative photoconductivity. The solid red line is the fitting curve with the single exponential decay function, which gives rise to the lifetime of 1.8 ps in the heterostructure with pump fluence of 24.8 $\mu J/cm^2$. (b) Transient THz transmission of monolayer $MoS_2$, no obvious signal is detected. (c) TA spectra of monolayer graphene and $MoS_2$ under photoexcitation of 780 nm, in which no clear signal is detected for both films.

**Section S3: Transient THz transmission of Gr/$MoS_2$ pumping with different photon energies.**

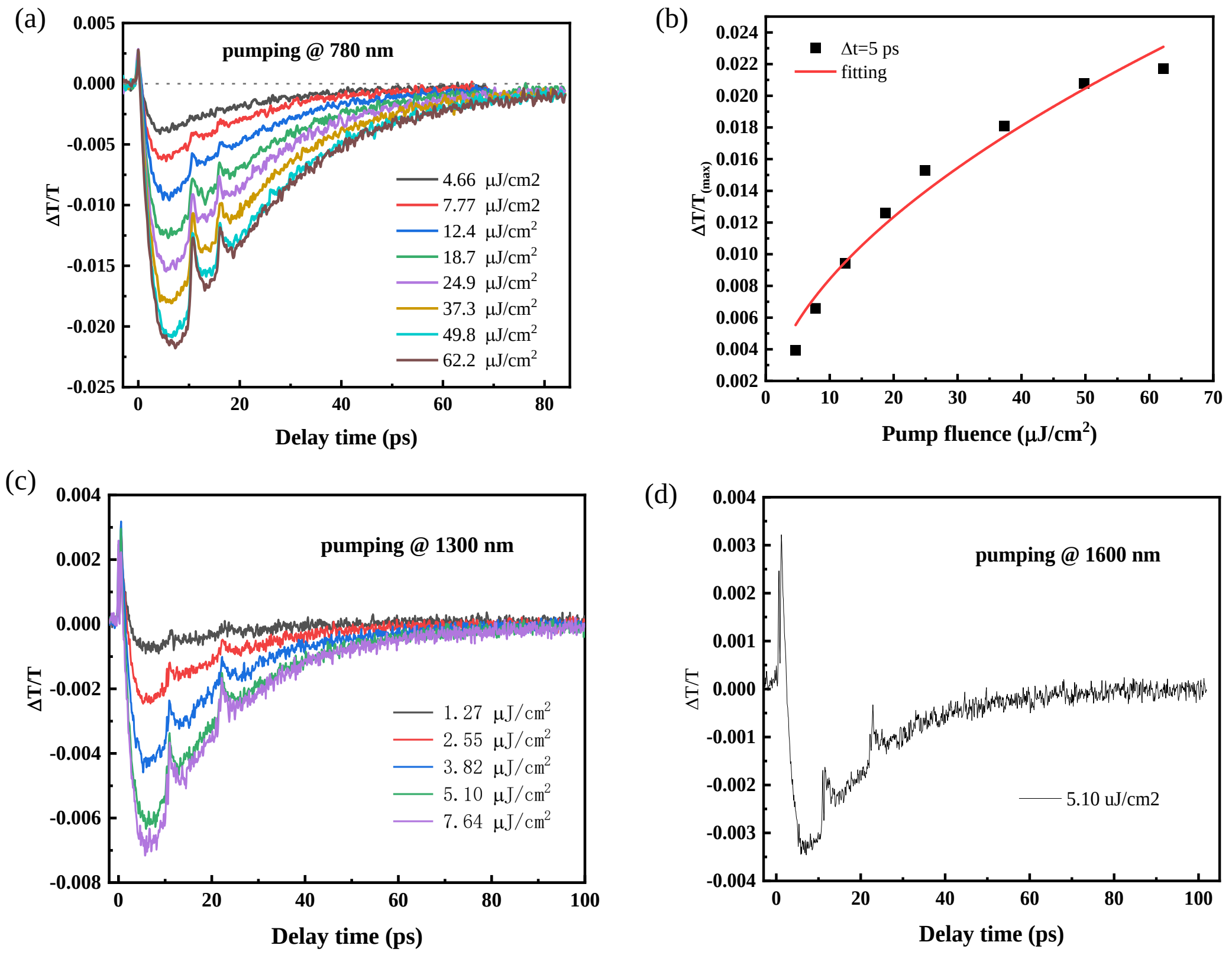

Figure S3. (a) Transient terahertz transmission of Gr/$MoS_2$ under 780 nm photoexcitation with several pump fluences. (b) The peak THz transmission $\Delta T/T_0$ at 5 ps (the minimum of $\Delta T/T_0$) in (a) with respect to pump fluence. Figures S3(c) and S3 (d) are the transient THz transmission of Gr/$MoS_2$ with pump wavelength of 1300 nm and 1600 nm, respectively.

Figure S3(a) shows the transient THz transmission of Gr/$MoS_2$ under 780 nm excitation with various pump fluences. We extracted the maximum positive photoconductivity, as well as its dependence on pump power, which is shown in Figure S3(b). It is obvious that sublinear rather than superlinear changes has been observed. In addition, we also have tested the transient THz transmission of Gr/$MoS_2$ at pump wavelength of 1300 nm and 1600 nm, as shown in S3 (c) and S3 (d) respectively. The transient THz transmission with pumping at 1300 and 1600 nm shows similar trace as that with 780 nm pump, which demonstrates the transfer of hot electrons from graphene to $MoS_2$ for all below A-exciton excitation.

## Section 4: Comparison of transient THz transmission excited with different photon energy

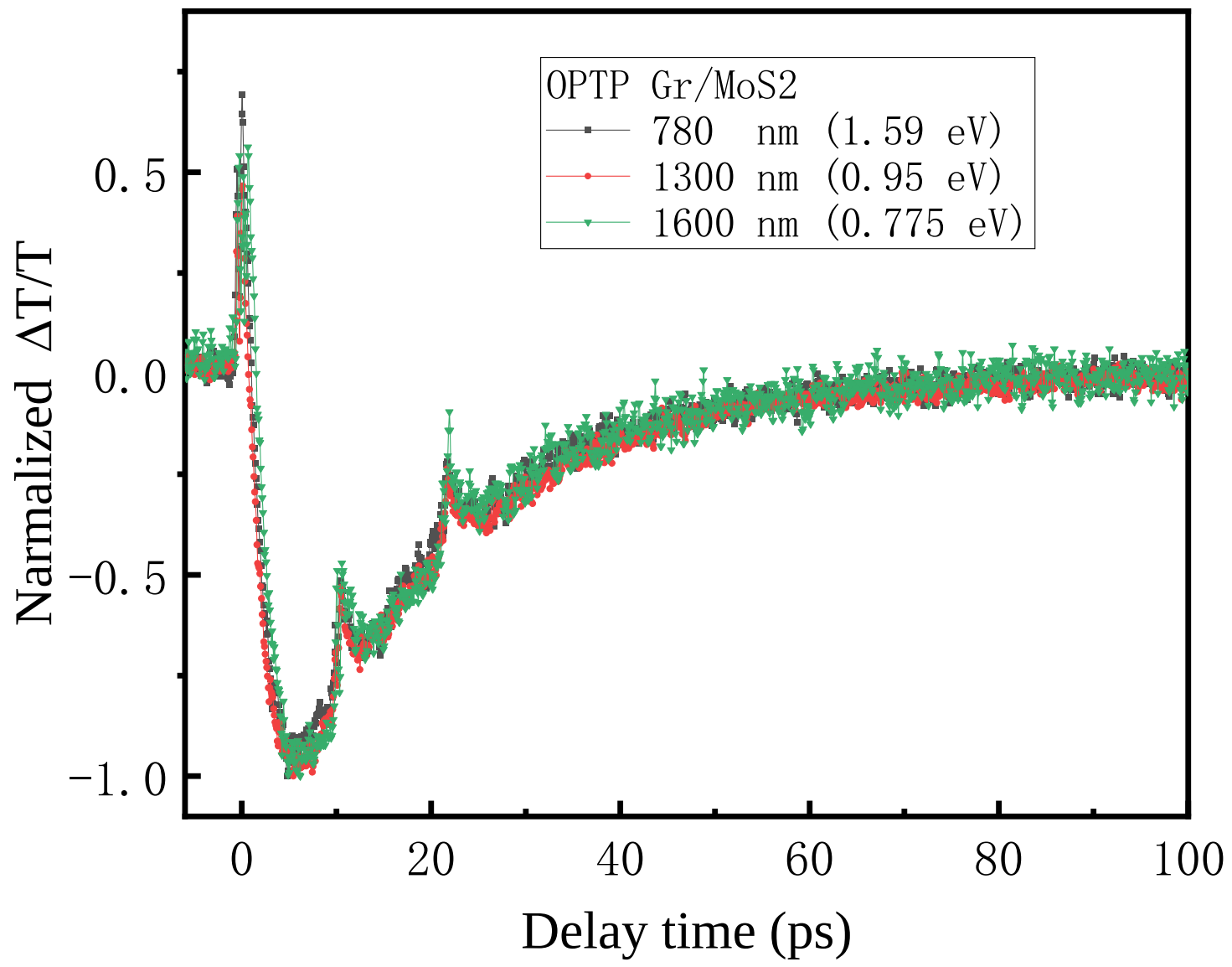

Figure S4. Normalized transient THz transmission of Gr/$MoS_2$ with photoexcitation with various photoenergy of 1.59 (780 nm), 0.95 (1300 nm) and 0.775 eV (1600 nm). It is clear the positive THz photoconductivity show identical relaxation under excitation of with different photon energy.

**Section S5. The photobleaching recovery signal of Gr/$MoS_2$ heterostructure under 780 nm excitation.**

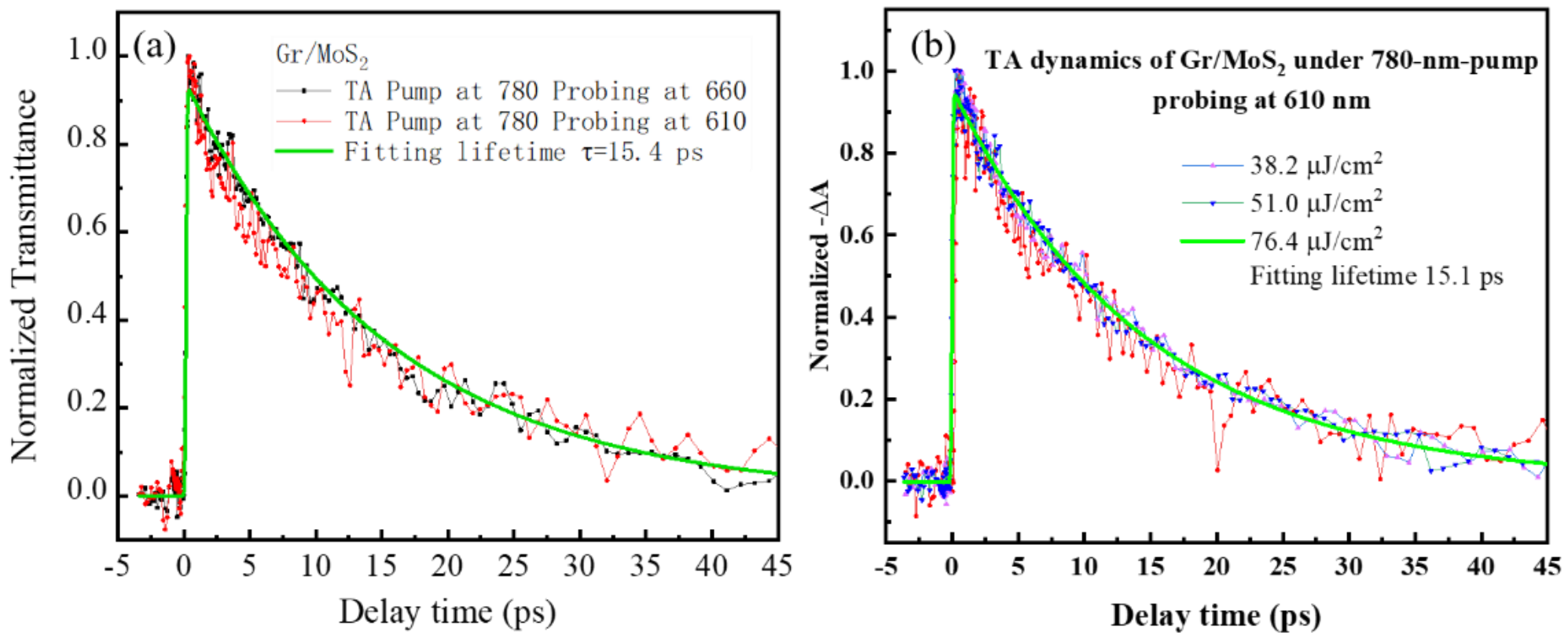

Figure S5. (a) Under excitation of 780 nm, the normalized transient carrier dynamics of Gr/$MoS_2$ at 660 nm (A-exciton) show basically the same recovery with various pump fluence. The green solid line is the single exponential fitting which produces time constant of 15.1 ps. (b) Transient dynamics at 660 nm (A-exciton, black) and 610 nm (B-exciton, red) under photoexcitation of 780 nm for Gr/$MoS_2$ heterostructure. The green line is the single exponential fitting which produces time constant of 15.4 ps.

**Section S6. The fitting process of $Gr/MoS_2$ transient THz transmission under below and above A-exciton optical excitation.**

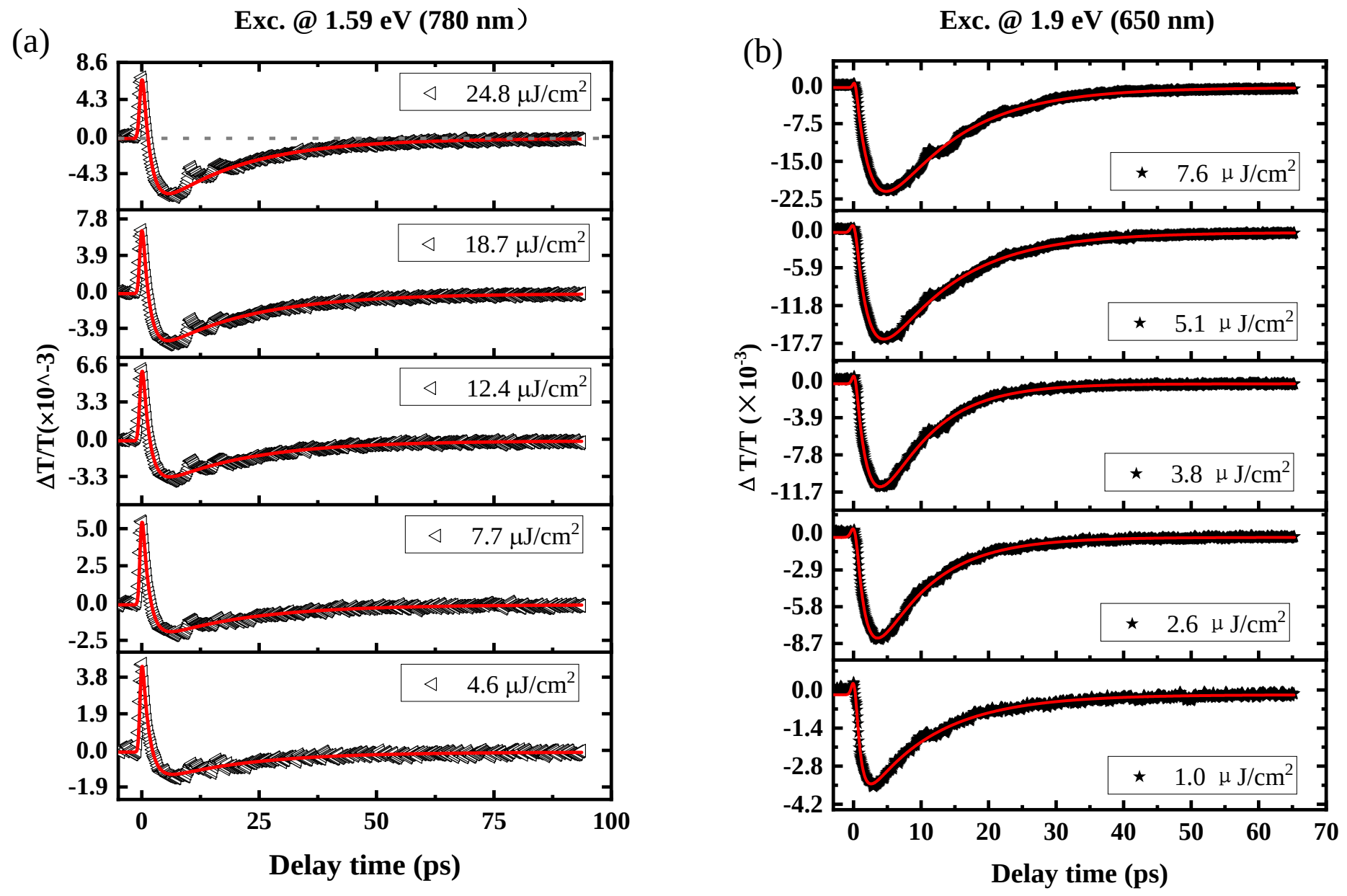

Figure S6. Transient THz response of $Gr/MoS_2$ under photoexcitation of 780 nm and 650 nm, the solid red line is numerical fitting of the double exponential decay function convolved with a 200 fs half-width Gaussian pulse. (a) and (b) are time-resolved spectra of 780 nm and 650 nm photoexcitation respectively.

Figures S6 (a) and (b) show the transient THz transmission of $Gr/MoS_2$ with pump wavelength of 780 nm (6(a)) for below A-exciton excitation and 650 nm (6(b)) for near resonance on A-exciton excitation. The convoluted exponential model with the Eq. S1 was employed to fit the ultrafast transient THz transmission response,

$$\frac{\Delta T}{T} = \left(\sum_i A_i \;\; exp^{-t/\tau_i}\right) \otimes \; IRF + B \qquad \text{S1}$$

where, $A_i$ is the amplitude of each decaying component and $\tau_i$ is the corresponding relaxation time. Function IRF is the instrument response function, and the last B is the background offset, from which we can obtain the exponential fitting function described by complementary error function erfc(x)=1-erf (x), as given below:

$$\frac{\Delta T}{T} = \sum_{i=1}^{2} A_i e^{\frac{\omega^2}{\tau_i^2} - \frac{t}{\tau_i}} \cdot erfc\left(\frac{\omega}{\tau_i} - \frac{t}{2\omega}\right) + \mathrm{B} \qquad \text{S2}$$

Origin program was used for numerical fitting of experimental data, and the results are shown in red solid lines in Figure S6.

## Section S7. TA spectra of $MoS_2$ monolayer along with the comparison between $MoS_2$ monolayer and Gr/$MoS_2$ heterostructure

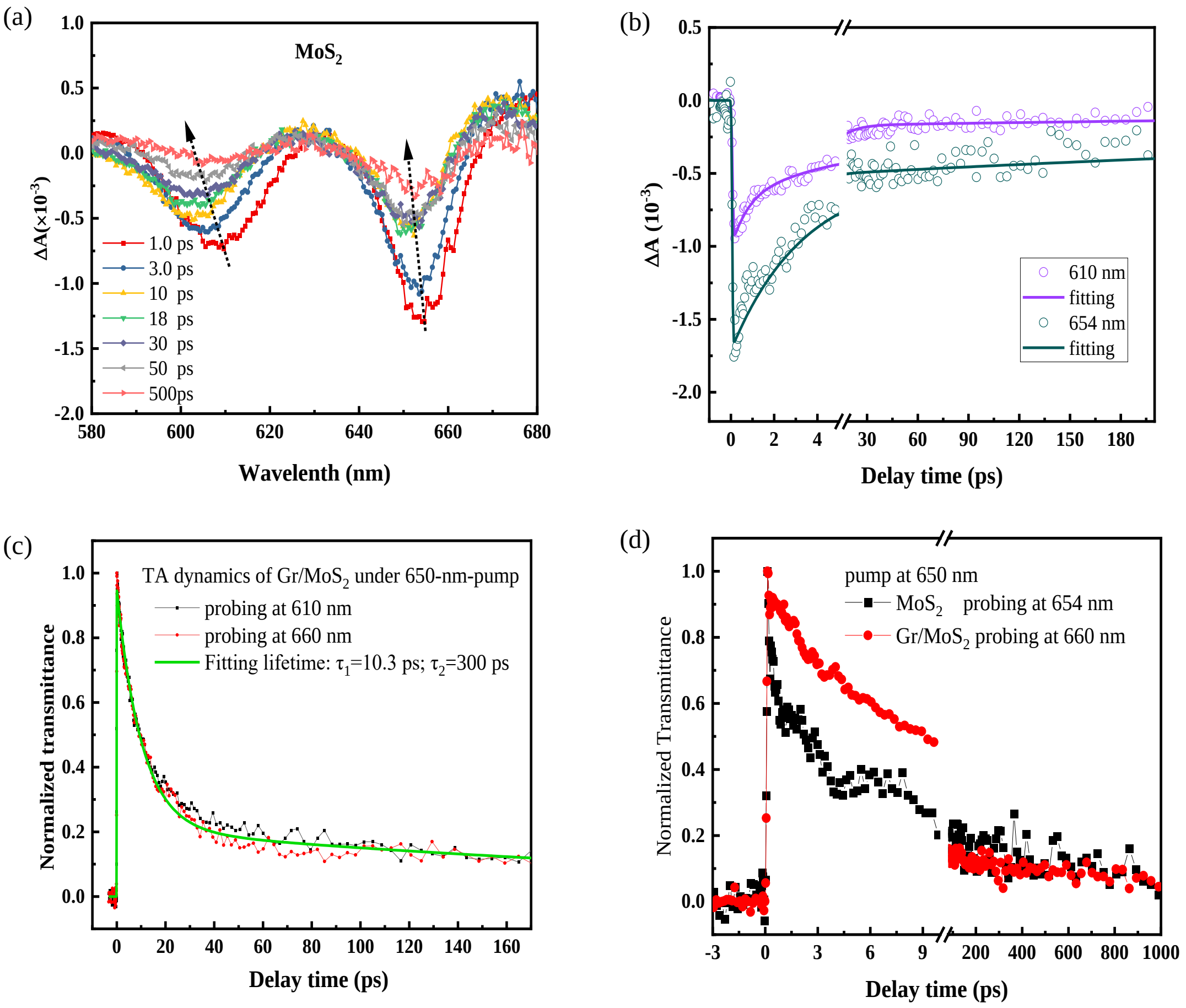

Figure S7. (a) TA spectra of $MoS_2$ monolayer with several selected delay times. (b) The dynamics of $MoS_2$ at probing wavelength of 610 nm (purple, B-exciton), 654 nm (dark blue, A-exciton). (c) Normalized transmittance of Gr/$MoS_2$ heterostructure under 650 nm pump with probing wavelength of 660 nm (black, B-exciton) and 610 nm (red, A-exciton) as well as biexponential fitting (green). (d) Normalized transient transmittance of $MoS_2$ monolayer (black) and Gr/$MoS_2$ heterostructure (red) under identical pump fluence at wavelength of 650 nm. The probing wavelengths are 654 nm for $MoS_2$ monolayer and 660 nm for Gr/$MoS_2$ heterostructure.

## Section S8. The photoconductivity versus frequency for Gr/$MoS_2$ heterostructure with various pump fluences at 780 nm.

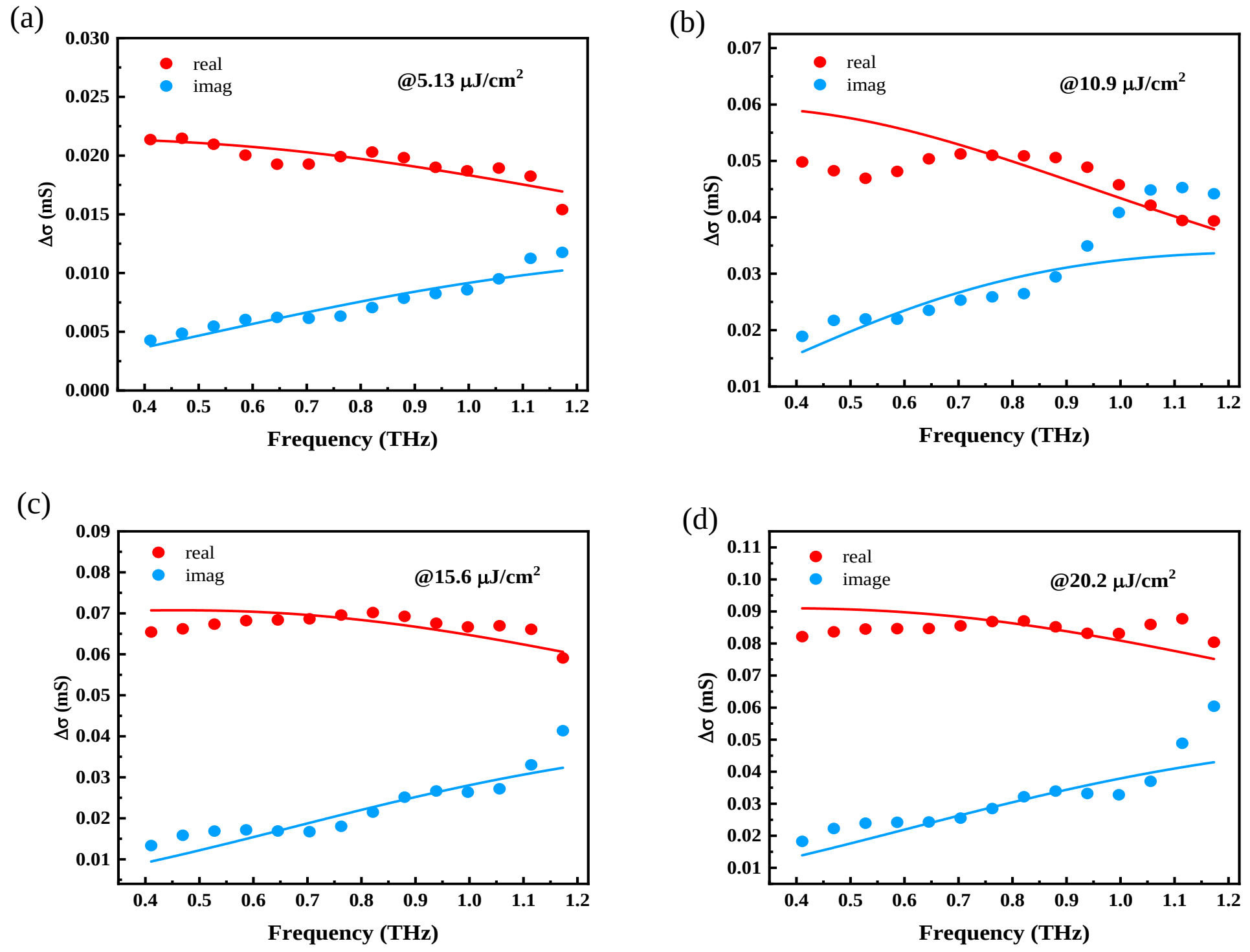

Figure S8. The photoconductivity dispersion of Gr/$MoS_2$ heterostructure with different pump fluence at 780 nm. The transient TDS spectra are collected at the delay time of 5 ps. Pump fluence from (a) to (d) are fixed at 5.13, 10.9, 15.6 and 20.2 $\mu J/cm^2$, respectively.

To calculate the number of transferred carriers in Gr/$MoS_2$ heterostructure after photoexcitation at 780 nm, the photoexcited THz time-domain spectra collected 5 ps were measured with a OPTP setup. The complex THz photoconductivity in frequency domain are obtained by Fourier transformation of time-domain spectra. Figure S8 shows the THz complex photoconductivity dispersion in the frequency range of 0.4 – 1.2 THz under the photoexcitation of 780 nm and various pump fluences.

To avoid the influence of the hot carrier, the time delay between the pumping optical and the THz pulse was set at 5 ps. The complex photoconductivity follows Drude-Smith model, as described in Eq. S3:

$$\Delta\tilde{\sigma}\ (\omega) = \frac{\varepsilon_0 \omega_p^2}{\Gamma - i\omega}\left[1 - \frac{c}{1 - i\omega/\Gamma}\right] \qquad \text{S3}$$

the first term on the right-hand side of Eq. S3 represents Drude conductivity, the second term is Smith correction, in which $\varepsilon_0$ and $\omega_p$ represent vacuum permittivity and plasma oscillation frequency respectively, $\Gamma$ corresponds to scattering rate, *c* represents the backscattering constant with the range of 0 to -1, *c*=0 denotes free charge carrier without experiencing any backscattering, while *c*=-1 presents the free charge carrier undergoing completely backscattering. As illustrated in Figure S8 above as well as in Figure 4. in the main text, the real and imaginary parts of the photoconductivity can be well reproduced Drude-Smith model with c approaching -0.3, indicating that the response of the photoconductivity comes from the contribution of free carriers.

## Section S9. Assessment of the number of carrier transfer in Gr/MoS$_2$ heterostructure with below A-exciton excitation.

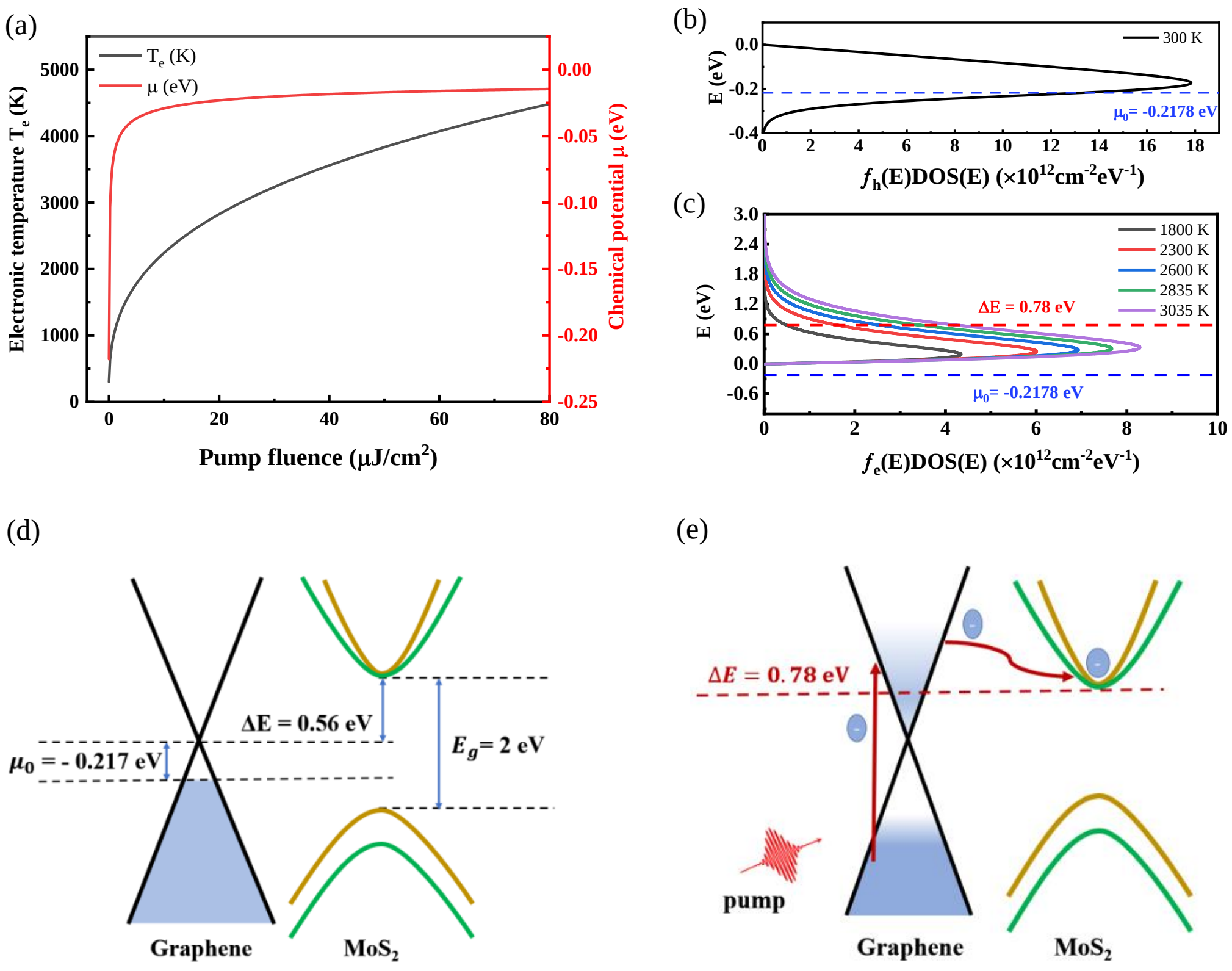

Figure S9. Theoretical evaluation of the number of hot electron transfer in Gr/MoS$_2$ heterostructures under 780 nm photoexcitation. (a) Electron temperature ($T_e$) and chemical potential ($\mu$) of graphene with respect to pump fluence. (b) The hole distribution of graphene layer at room temperature of 300K. (c) Fermi-Dirac distribution of electrons at several selected temperatures. (d) Diagram of energy band gap for Gr/MoS$_2$ at room temperature before optical excitation. (e) Schematics of hot electrons transfer from graphene cross the interfacial barrier and injected into MoS$_2$ layer under 780 nm excitation.

Consider a p-type graphene, the pump fluence dependent electron temperature $T_e$ can be described by Eq. S4 under photoexcitation.[1-2]

$$T_e = T_L\left(1 + \frac{3\gamma F}{\beta T_L^3}\right)^{1/3} \qquad \text{S4}$$

in which, $T_L$ represents the initial lattice temperature of 300 K. $\gamma$ is the ratio of the energy captured by graphene electron to the incident photon, with $\gamma$=70%×2.3%, according to previous work, 70% of incident photon energy can be applied, whereas graphene absorbs about 2.3% of incident photons.[3-4] F is the energy density (with unit of $J/cm^2$) of the incident photons. $\beta = \frac{18\xi(3)K_B^3}{(\pi\hbar v_F)^2}$, $\xi(3) = 1.202$, where $K_B$ is Boltzmann constant, $v_F$ is graphene Fermi velocity ($v_F = 1.0 \times 10^{\wedge}6$), and $\hbar$ is reduced Planck constant. Furthermore, the chemical potential of graphene with respect to temperature can be calculated according to Eq. S5.[1,5] In short, electron temperature and chemical potential of graphene as functions of pump fluence is plotted in Figure S9(a).

$$\mu = \frac{E_F^2}{4\ln(2)K_B T_e} \qquad \text{S5}$$

Before photoexcitation, the static conductivity ($\sigma_0$) of graphene is contributed by holes of the p-type graphene. The initial carrier density of graphene at room temperature is calculated by Eq. S6:

$$n_0 = \int_{\mu_0}^{0} (1 - f_e(E)) DOS(E) d(E) \qquad \text{S6}$$

where $f_e(E) = \frac{1}{e^{(E-\mu)/K_B T_e}+1}$ is the Fermi-Dirac distribution of electrons, $DOS(E) = \frac{2E}{\pi(\hbar vF)2}$ is the graphene density of states. Figure S9(b) shows the calculated results at temperature of $T_e = 300\ K$, and $\mu_0$ is the initial chemical potential without photoexcitation.

The electron temperature in graphene can be heated from 1800 K to 3035 K depending on the pump fluence given in our experimental data. At these temperatures, the distribution of thermal electrons is shown in Figure S9(c). From Eq. S7, we can obtain the number of electrons injected from graphene into $MoS_2$. It is assumed that all the hot electrons above the interfacial barrier are transferred into $MoS_2$, as illustrated in Figure S9(e).

$$\Delta n = \int_{\Delta E}^{\infty} f_e(E) DOS(E) d(E) \qquad \text{S7}$$

The ratio of photoconductivity ($\Delta\sigma$) to static conductivity ($\sigma_0$) in Gr/$MoS_2$ heterostructure is then given by Eq S8:

$$\frac{\Delta\sigma}{\sigma_0} = \frac{\Delta n}{n_0} = \frac{\int_{\Delta E}^{\infty} f_e(E) DOS(E) d(E)}{\int_{\mu_0}^{0} \left(1 - f_e(E)\right) DOS(E) d(E)} \tag{S8}$$

As the measured value $\Delta\sigma/\sigma_0$ approximately equal to the $\Delta n/n_0$, therefore, we can evaluate how much carrier transfer takes place based on the measured quanta of $\Delta\sigma/\sigma_0$.